\documentclass[]{aa}

\usepackage{graphicx}
\usepackage{subfig}

\usepackage[export]{adjustbox}
\usepackage{placeins}

\usepackage[pdfstartview=XYZ,
bookmarks=true,
colorlinks=true,
linkcolor=blue,
urlcolor=blue,
citecolor=blue,
pdftex,
bookmarks=true,
linktocpage=true, 
hyperindex=true
]{hyperref}

\usepackage{orcidlink}

\usepackage{txfonts}
\usepackage{hyperref}
\usepackage{natbib}

\linenumbers

\begin{document}

   \title{Non-thermal GeV emission from the Nereides nebula: confirming the nature of the supernova remnant G$107.7-5.1$}


   \author{Miguel Araya\,\orcidlink{0000-0002-0595-9267}}

   \institute{Escuela de F\'isica, Universidad de Costa Rica \\
              Montes de Oca, San Jos\'e, 11501-2060, Costa Rica\\
              \email{miguel.araya@ucr.ac.cr}}

   \date{Accepted 20/09/2024}

 
  \abstract
   {Recently the Nereides nebula was discovered through deep optical emission line observations and was classified as a supernova remnant (SNR) candidate, G$107.7-5.1$.}
   {Since very little is known about this SNR we look at several archival data sets to better understand the environment and properties of the object.}
   {We present a detailed analysis of the gamma-ray emission detected by the \textit{Fermi} Large Area Telescope in the region of the nebula. A model of the non-thermal emission is presented which allows us to derive the particle distribution responsible for the gamma rays. We also use molecular gas and atomic hydrogen observations to try to constrain the source age and distance.}
   {An extended ($\sim 2^\circ$) GeV source coincident with the location of the nebula is found. The non-thermal emission has a hard spectrum and is detected up to $\sim 100$~GeV, confirming the SNR nature of this object. The GeV properties of G$107.7-5.1$ are similar to those of other SNRs such as G$150.3+4.5$, and likely expands in a relatively low-density medium. The Nereides nebula is one more example of a growing population of dim SNRs detected at high energies. A simple leptonic model is able to account for the gamma-ray emission. Standard SNR evolutionary models constrain the age to be in the $10-50$~kyr range, which is consistent with estimates of the maximum particle energy obtained from GeV observations. However, more detailed observations of the source should be carried out to better understand its properties.}
   {}


   \maketitle
%

\section{Introduction} \label{sec:intro}
Supernova remnants (SNRs) play an important role in the interstellar medium, modifying its composition and temperature, affecting star formation activity and accelerating cosmic rays to high energies. Studying the properties of SNRs can also help understand the progenitors and their evolution. Some $\sim 300$ Galactic SNRs are known today \citep{2019JApA...40...36G,2012AdSpR..49.1313F}, most of which were discovered by radio surveys. This number represents a mismatch between the expected \citep[see, e.g.,][]{1991ApJ...378...93L} and the observed number of SNRs in our Galaxy. The discrepancy may be partly explained due to selection effects such as the sensitivity limits of surveys.

Recently, more SNR and SNR candidates have been discovered using X-ray and optical observations, in some cases accompanied by non-thermal gamma-ray detections \citep[e.g.,][]{2017A&A...605A..58A,2020MNRAS.498.5194F,2020MNRAS.493.2188G,2021A&A...648A..30B,2021MNRAS.507..971C,2022MNRAS.510.2920A,2022A&A...667A..71A,2023AJ....166..149F,2023MNRAS.521.5536K,2024A&A...683A.107Z,2024ApJS..272...36F}. In particular, new SNRs have been found at varying heights above the Galactic plane. Studying the properties of SNRs outside the Galactic plane such as G$181.1+9.5$ \citep{2017A&A...597A.116K} and G$288.8-6.3$ \citep{2023AJ....166..149F} can help understand the environment and magnetic field in the outer disk and halo as well as SNR evolution in these locations \citep[e.g.,][]{Shelton_1998}.

Deep all-sky surveys such as the one carried out at GeV energies by the Large Area Telescope (LAT) onboard the \textit{Fermi} satellite \citep{2009ApJ...697.1071A} have provided a valuable tool to search for SNR candidates which often appear as extended sources. For many unidentified extended sources in the LAT catalogs there is no known counterpart at lower energies \citep{Abdollahi_2020}. They could potentially be associated with SNRs or pulsar wind nebulae and serve to confirm the SNR nature of newly found candidates at lower energies. SNRs can produce gamma rays through the interactions of the high-energy cosmic ray electrons or protons accelerated in their shocks with ambient photon fields or gas, respectively.

In this work we investigate the properties of the recently discovered SNR G$107.7-5.1$ \citep[the Nereides nebula,][]{2024ApJS..272...36F}. Optical observations revealed G$107.7-5.1$ as a very faint shell of filaments nearly 3 degrees in size which, together with its relatively high Galactic latitude \citep{2024ApJS..272...36F} might explain why the source has not been found in radio surveys. We confirm the SNR nature of the object with the detection of a GeV counterpart in \textit{Fermi}-LAT data. This makes the Nereides nebula the 12th confirmed SNR at relatively high Galactic latitudes that is detected by \textit{Fermi}, including several radio-dim SNRs listed by \cite{2024A&A...684A.150B}.

An extended GeV source with a hard spectrum ($\frac{dN}{dE} \propto E^{-\Gamma}$ with $\Gamma = 1.95\pm 0.08_{\mbox{\tiny stat}} \pm 0.15_{\mbox{\tiny sys}}$), FHES~J$2304.0+5406$, was discovered by \cite{2018ApJS..237...32A} at the location of the now known SNR G$107.7-5.1$. These authors proposed that FHES~J$2304.0+5406$ was likely associated to an SNR or a pulsar wind nebula. The morphology of this high-energy source was modeled with a Gaussian template with a 68\%-containment radius of $1.58 \pm 0.35_{\mbox{\tiny stat}} \pm 0.17_{\mbox{\tiny sys}}^\circ$. In the latest incremental version of the fourth \textit{Fermi}-LAT catalog \citep[4FGL-DR4,][]{2023arXiv230712546B}, the corresponding source is labeled 4FGL~J$2304.0+5406$e having the same morphology as FHES~J$2304.0+5406$ and a spectrum described by a log-parabola, $\frac{dN}{dE} = N_0\left(\frac{E}{E_b}\right)^{-\alpha -\beta\,\mbox{ln}\,(E/E_b)}$, where $\alpha = 1.72 \pm 0.08$ and $\beta = 0.08\pm 0.04$ and $E_b$ is a fixed scale parameter. The significance of ``spectral curvature'' (i.e., the deviation of the source spectrum from a simple power-law), however, was reported to be at the $2.6\sigma$ level.

\section{Data}
\subsection{\textit{Fermi}-LAT observations}
We downloaded LAT Pass 8 observations\footnote{See \url{https://fermi.gsfc.nasa.gov/ssc/data/access/}} from the beginning of the mission, 2008 August, to 2024 March, containing events reconstructed within $25^\circ$ of the coordinates RA\,$=346.0^\circ$, Dec\,$=54.6^\circ$ in the energy range 0.1--500~GeV. Front and back-converted \texttt{SOURCE} class events were used for the analysis (\texttt{evclass}=128, \texttt{evtype}=3) which was done with the software \texttt{fermitools} version 2.2.0 \footnote{See \url{https://github.com/fermi-lat/Fermitools-conda/}} through the \texttt{fermipy} package version 1.2.0 \citep{2017ICRC...35..824W}. We binned the data with a spatial scale of $0.05^\circ$ per pixel and ten bins in energy for exposure calculations. The analysis used the publicly-available response functions appropriate for the data set, \texttt{P8R3\_SOURCE\_V3}.

The data analysis relies on the maximum likelihood technique \citep{1996ApJ...461..396M}. For a given spectral and morphological model of a source its emission is convolved with the response functions to predict the number of counts in the spatial and energy bins. A fit of the free parameters is carried out to maximize the probability for the model to explain the data in each bin. The detection significance of a new source having one additional free parameter, for example, can be calculated as the square root of the test statistic (TS), defined as TS$=-2 \,\mbox{log}\,(\mathcal{L}_0/\mathcal{L})$, with $\mathcal{L}$
and $\mathcal{L}_0$ the maximum likelihood functions for a model with the new source and for the model without this additional source (the null hypothesis) respectively. In all the models the diffuse Galactic emission is described by the file \texttt{gll\_iem\_v07.fits}, and the isotropic emission and residual cosmic-ray background is given by \texttt{iso\_P8R3\_SOURCE\_V3\_v1.txt}, which are provided with the \texttt{fermitools}. As recommended by the LAT team the energy dispersion correction was applied to all sources except for the isotropic component.

\subsubsection{Morphology of the GeV emission}
We first carried out a morphological analysis of the gamma-ray emission in the region of G$107.7-5.1$. For this purpose, we took advantage of the improved resolution at higher energies and only used events with energies above 10~GeV. To avoid gamma-ray contamination from cosmic-ray interactions in the Earth's atmosphere we set a maximum event zenith angle of $105^\circ$ \citep{Abdollahi_2020}. In this step we used a region of interest with a radius of $15^\circ$ around RA\,$=346.0^\circ$, Dec\,$=54.6^\circ$ and included in the model the cataloged sources from 4FGL-DR4 found within $20^\circ$ of the center. To optimize the model initially we carried out a preliminary fit of the spectral normalizations of the 4FGL-DR4 sources located within $10^\circ$ of the center, as well as the spectral indices and normalizations of the sources located within $5^\circ$ of the center of the region. We set the normalizations of the diffuse components free in all fits.

We searched for potentially new point sources in the region to improve the model of the background using the \texttt{fermipy} function \texttt{find\_sources}. This algorithm searches for peaks in the TS map obtained when subtracting the cataloged sources. In this case we only set free the normalizations of the model sources found within $3^\circ$ from the center. We added new source candidates to the model if their calculated $\sqrt{\mbox{TS}}$ was above a threshold of 4. The spectral function used for the point sources was a simple power-law. After adding the new candidate sources to the model we set the parameters of the sources free as before (the source normalizations for sources within $10^\circ$ of the center, and the spectral indices and normalizations for sources within $5^\circ$ of the center) and performed a new fit.

We compared two alternative models for the emission from G$107.7-5.1$: the 2D gaussian template used in the 4FGL-DR4 catalog and a uniform 2D disk. After removing the source 4FGL~J$2304.0+5406$e from the model we searched for the disk radius ($r$) and center location which maximize the likelihood (again with only the normalizations of the model sources located within $3^\circ$ from the center free to vary) resulting in the values $r=1.02\pm 0.03^\circ$, RA$=346.14\pm 0.04^\circ$, Dec$=54.31\pm 0.05^\circ$ ($1\sigma$ statistical uncertainties quoted).

After freeing the background sources again as explained before, we performed two independent fits using the disk and Gaussian templates, assuming simple power-laws for the spectral shape with free index and normalization. The fit using the disk results in a slightly higher likelihood value with TS$=72.5$, compared to the fit with the Gaussian which gives TS$=68.2$. We thus adopted the disk template to model the GeV emission from the SNR and used the Gaussian morphology to estimate systematic errors in the spectral parameters resulting from changing the morphology (see below).

\begin{figure}
\hspace*{-4cm}
\includegraphics[width=1.8\columnwidth]{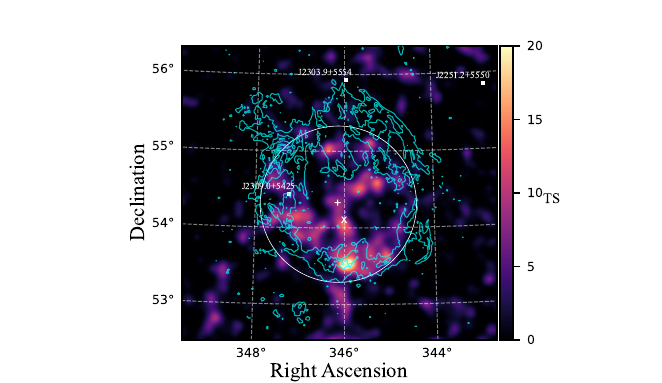}
\caption{TS map for events with energies above 2~GeV showing the emission from G$107.7-5.1$. The circle represents the uniform disk obtained in this work to model the morphology of the gamma-ray emission. The contours show the optical emission of the SNR shell taken from \cite{2024ApJS..272...36F}. The marks \tt{+} and \tt{x} indicate the center locations of the disk and the extended 4FGL source, respectively, while the boxes show the locations of the other 4FGL-DR4 sources in the region.\label{fig1}}
\end{figure}

In order to obtain a map of the GeV emission with improved statistics, the procedure described above for optimizing the model was repeated for events with energies above 2~GeV. Our model provides a satisfactory fit to the background sources (see Appendix \ref{residuals}). The best-fit disk found in this section is removed from the model and a TS map is calculated by optimizing the spectral normalization of a putative point source that is moved in a grid of positions, and calculating its TS in each fit. We used a spectrum described by a power-law for the point source with a fixed index of 2. Fig. \ref{fig1} shows the resulting TS map. The extension and location of the best-fit GeV disk is shown in the figure as well as the center location of the 4FGL Gaussian model and the locations of background 4FGL-DR4 sources. Our disk model matches the location of the SNR shell while using the catalogued source 4FGL~J$2304.0+5406$e does not improve the fit and it is more extended than the SNR. The extension of this source was determined by \cite{2018ApJS..237...32A} using events in the energy range 1~GeV--1~TeV, while we used events above 10~GeV having a much better resolution.

The point source 4FGL~J$2309.0+5425$ is seen in the direction of the shell of the SNR and according to \cite{2023arXiv230712546B} could be related to the X-ray source 1RXS~J$230852.2+542559$. The source 4FGL~J$2303.9+5554$ is seen just north of the SNR shell and was classified as a blazar candidate by \cite{2021ApJ...923...75K}. Both of these sources are clearly point-like in LAT data and thus likely unrelated to the shell emission (see Appendix \ref{10GeVmap}), and therefore they were kept in the background model. Fig. \ref{fig1} also shows contours from the optical [O III] emission of the SNR, obtained from \cite{2024ApJS..272...36F}. [O III] emission is associated with ionization, resulting for example from high-velocity shocks, and can be used to trace the remnant's actual morphology. A consistency is seen between the extension and location of the gamma-ray emission and the SNR.

\subsubsection{Gamma-ray spectrum}
We collected all events in the energy range 0.1--500~GeV to study the high-energy spectrum of G$107.7-5.1$. We used similar filtering parameters as in the previous section except for the maximum zenith angle, which was set to $90^\circ$. To account for the larger point spread function of low energy events we also increased the radius of the region of interest to $20^\circ$ and added cataloged sources located up to a distance of $25^\circ$ from the center of the region. We modeled the SNR using the disk obtained in the previous section and searched for new point source candidates to improve the background model (see Appendix \ref{residuals}). After optimizing the spectral parameters of sources following the procedure of the previous section we produced a spectral energy distribution (SED) of G$107.7-5.1$ dividing the data in 11 energy bins equally spaced in a logarithmic scale. In each bin, the spectrum of the SNR is modeled using a power-law with a fixed spectral index of the form  $\frac{dN}{dE} = N_0 \left( \frac{E}{1\,\mbox{\tiny GeV}}\right)^{-2}$ and the normalization $N_0$ is fit. We did not significantly detect the source in any bin below $\sim 1$~GeV or above $\sim 100$~GeV and derived 95\%-confidence level upper limits (UL) for those bins, as can be seen in Table \ref{table1_sed}.

\begin{table}
\begin{center}
\caption{SED measurements of SNR G$107.7-5.1$}
\label{table1_sed}
\begin{tabular}{cc}
\hline\hline
Energy (GeV) & $E^2\frac{dN}{dE}$ ($10^{-12}$~erg~cm$^{-2}$~s$^{-1}$) \\
\hline
0.15 & 0.45\tablefootmark{a} \\
0.32 & 0.13\tablefootmark{a} \\
0.69 & 0.71\tablefootmark{a} \\
1.50 & $0.68\pm 0.32$ \\
3.26 & $0.86\pm 0.39$ \\
7.07 & $1.82\pm 0.39$ \\
15.3 & $1.63\pm 0.54$ \\   
33.3 & $3.28\pm 0.66$ \\
72.2 & $3.26\pm 0.95$ \\
156.5 & 2.77\tablefootmark{a} \\
339.5 & 4.55\tablefootmark{a} \\ 
\hline
\end{tabular}
\tablefoottext{a}{95\%-confidence level UL}
\end{center}
\end{table}

With the goal of constraining the spectral shape of the SNR we compared two fits in the energy range 0.1--500~GeV using a simple power-law and a log-parabola. The log-parabola resulted in a better fit to the data with a difference in TS values $\Delta\,$TS$=23$. For an additional degree of freedom in the model this corresponds to an improvement at the $4.8\sigma$ level, which demonstrates that the source spectrum is significantly curved. This is consistent with the results shown in the 4FGL catalog for the source 4FGL~J$2304.0+5406$e, where its spectrum is modeled with a log-parabola, although the reported significance of curvature is lower. Adopting the morphology of the 4FGL catalog (2D Gaussian), however, does not provide an improvement in the fit, resulting in TS $= 100.3$, compared to TS $=98.8$ obtained with the disk template found in the previous section. This result is consistent with the morphological study in this work.

A fit to the spectral function $\frac{dN}{dE} = N_0\left(\frac{E}{E_b}\right)^{-\alpha -\beta\,\mbox{ln}\,(E/E_b)}$ in the energy range 0.1--500~GeV produced the values $N_0 = (8.5 \pm 0.2 ^{+8.0}_{-0.2}) \times 10^{-15}$~MeV$^{-1}$~s$^{-1}$~cm$^{-2}$, $\alpha = 1.67\pm 0.01 \pm0.12$ and $\beta = 0.22\pm 0.01 ^{+0.09}_{-0.04}$ (the first errors are statistical while the second are systematic) for a fixed parameter $E_b=13$~GeV. We considered two factors that affect the source modeling to estimate the systematic errors in the parameters, the source morphology and the uncertainty in the effective area of the LAT. For the first we calculated the difference between parameter values obtained using the 2D Gaussian template from the 4FGL catalog and those found with the disk template. For the second effect we used a set of bracketing response functions following \cite{Ackermann_2012}, fixing the value of the pivot energy at $E_b = 13$~GeV for which the propagated error in the differential flux was minimal, to estimate the differences in the parameters obtained using the alternative and the nominal effective area models. For the combined systematic uncertainty, we added the errors from both effects in quadrature. The cause of the large error in the spectral normalization $N_0$ is the alternative Gaussian template which predicts about twice the flux for the source.

The gamma-ray energy flux of the source integrated above 1~GeV is $\sim 5.8\times 10^{-6}$~MeV~cm$^{-2}$~s$^{-1}$ corresponding to a luminosity of $\sim 1.1\times 10^{33}\,d_{\mbox{\tiny kpc}}^2$~erg~s$^{-1}$, where $d_{\mbox{\tiny kpc}}$ is the source distance in kpc.

\subsection{GeV emission model}\label{leptonic}
Non-thermal emission in SNRs can be produced by the relativistic particles accelerated at the shock. High-energy electrons are responsible for the synchrotron emission in the radio that is characteristic of many SNRs. These same electrons can produce gamma rays through inverse Compton scattering of low energy ambient photons, or through bremsstrahlung emission resulting from interactions with ambient nuclei if the densities are high enough. These mechanisms are thus leptonic. In the hadronic scenario, the high-energy nuclei can produce $\pi$ mesons in collisions with ambient matter if the density is high enough. The resulting neutral pions decay into gamma rays which can be detected. They are usually observed in regions where SNRs interact with molecular clouds.

\begin{figure}
\hspace*{-0.5cm}
\includegraphics[width=1.2\linewidth]{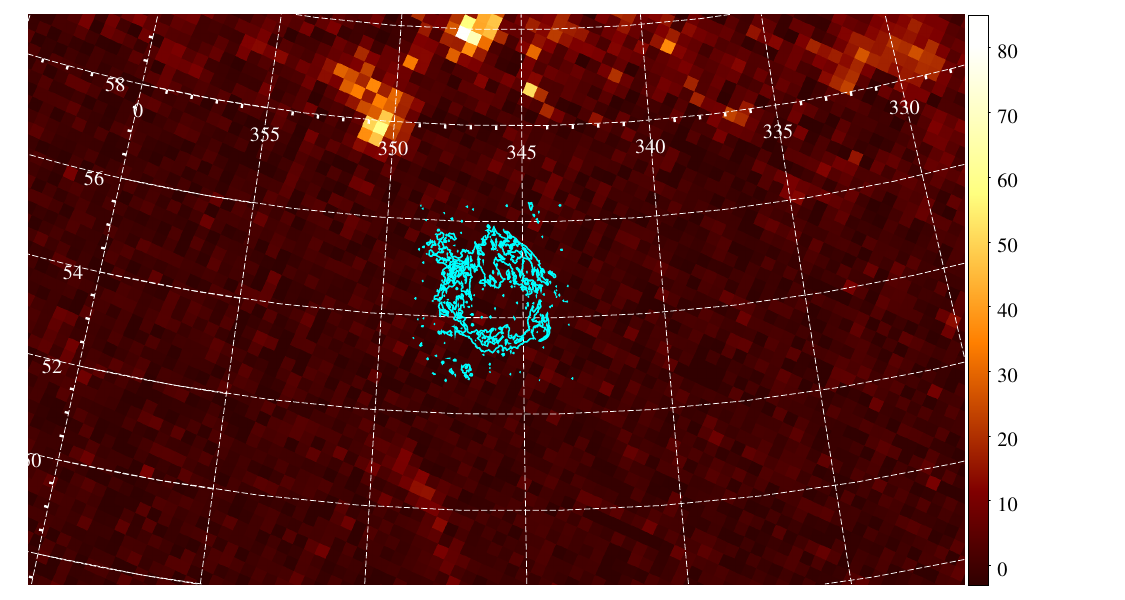}
\caption{Velocity-integrated CO ($J=1-0$) emission line from \cite{2001ApJ...547..792D} with the optical contours of the SNR overlaid (cyan, same as Fig. \ref{fig1}). Map units are K~km~s$^{-1}$ and grid coordinates are RA and Dec ($^\circ$).\label{fig2}}
\end{figure}

The rotational $J = 1-0$ transition of the carbon monoxide molecule has been widely used as a tracer of molecular hydrogen in astronomy. We downloaded data from the CO survey by \cite{2001ApJ...547..792D} (individual survey DHT18) to look for emission line signals in the direction of G$107.7-5.1$. No significant emission is apparent at any velocities, and Fig. \ref{fig2} shows the integrated emission in a region of the sky around the location of the SNR.

Although more detailed observations should be carried out, it is possible as these data suggest that G$107.7-5.1$ is evolving in a low-density environment (see Section \ref{discussion} for an estimate of the ambient densities). This seems reasonable given the location of the SNR $5^\circ$ below the Galactic plane. In fact, as shown in Fig. \ref{fig1}, the gamma-ray emission is more prominent in the south of the SNR, which is the region that is farthest from the Galactic plane.

We applied a simple one-zone leptonic model to account for the GeV flux from G$107.7-5.1$ to constrain the parameters of the particle distribution responsible for the emission. Given the likely low density environment of the SNR we do not consider a hadronic origin. We compared two functions for the underlying lepton distribution, a simple power-law (PL) of the form $\frac{dN}{dE_e} = a\, E_e^{-s}$, and a power-law with an exponential cutoff (EC), $\frac{dN}{dE_e} = a\, E_e^{-s}\,\mbox{e}^{-E_e/E_c}$, where $E_e$ is the electron energy, $s$ and $E_c$ are the spectral index and energy cutoff, respectively, and $a$ the normalization. We fit the observed fluxes from Table \ref{table1_sed} with an inverse Compton model using the \texttt{NAIMA} package \citep{naima}. The IC calculations are implemented from \cite{2014ApJ...783..100K}.

For the seed photon fields we included the cosmic microwave background radiation (CMB) and the local far infrared (FIR) and optical (NIR) fields. While the first is modeled as a black body, the other two are modeled as black bodies with a dilution factor, having the following densities and temperatures: FIR: $0.2$~eV~cm$^{-3}$ and $30$~K, NIR: $0.3$~eV~cm$^{-3}$ and $3000$~K. These values were estimated by \cite{2018A&A...617A..78T} using a model from \cite{2017MNRAS.470.2539P}.

Comparing the Bayesian Information Criterion (BIC) from the fits we obtained BIC$_{\mbox{\tiny PL}}-\,$BIC$_{\mbox{\tiny EC}}=2.07$ and thus the particle distribution having an exponential cutoff is a better model.

Although with the observations analyzed in this work we cannot constrain the synchrotron flux level from the source, modeling the GeV observations under the IC scenario allows us to determine the present particle slope and cutoff energy, however future observations, particularly at the highest energies, should be carried out to better constrain $E_c$. The resulting values are $s=1.98_{-0.11}^{+0.18}$ and $E_c = 4.2_{-1.2}^{+3.9}$~TeV, respectively. Assuming a source distance of 1~kpc, the total energy required in the particles is $8.6\times 10^{46}$~erg. If we fit the GeV spectrum with a particle distribution in the form of a broken power-law, the values obtained for the spectral indices before and after the break are $2.06_{-0.42}^{0.31}$, $2.98_{-0.50}^{+0.39}$, respectively, with a break energy of $4.0_{-1.8}^{+2.8}$~TeV, however, the fit does not represent an improvement with respect to the more simple exponential cutoff model, with an increase of $\sim3.5$ in the value of the BIC.

\begin{figure}
\includegraphics[width=1.1\linewidth]{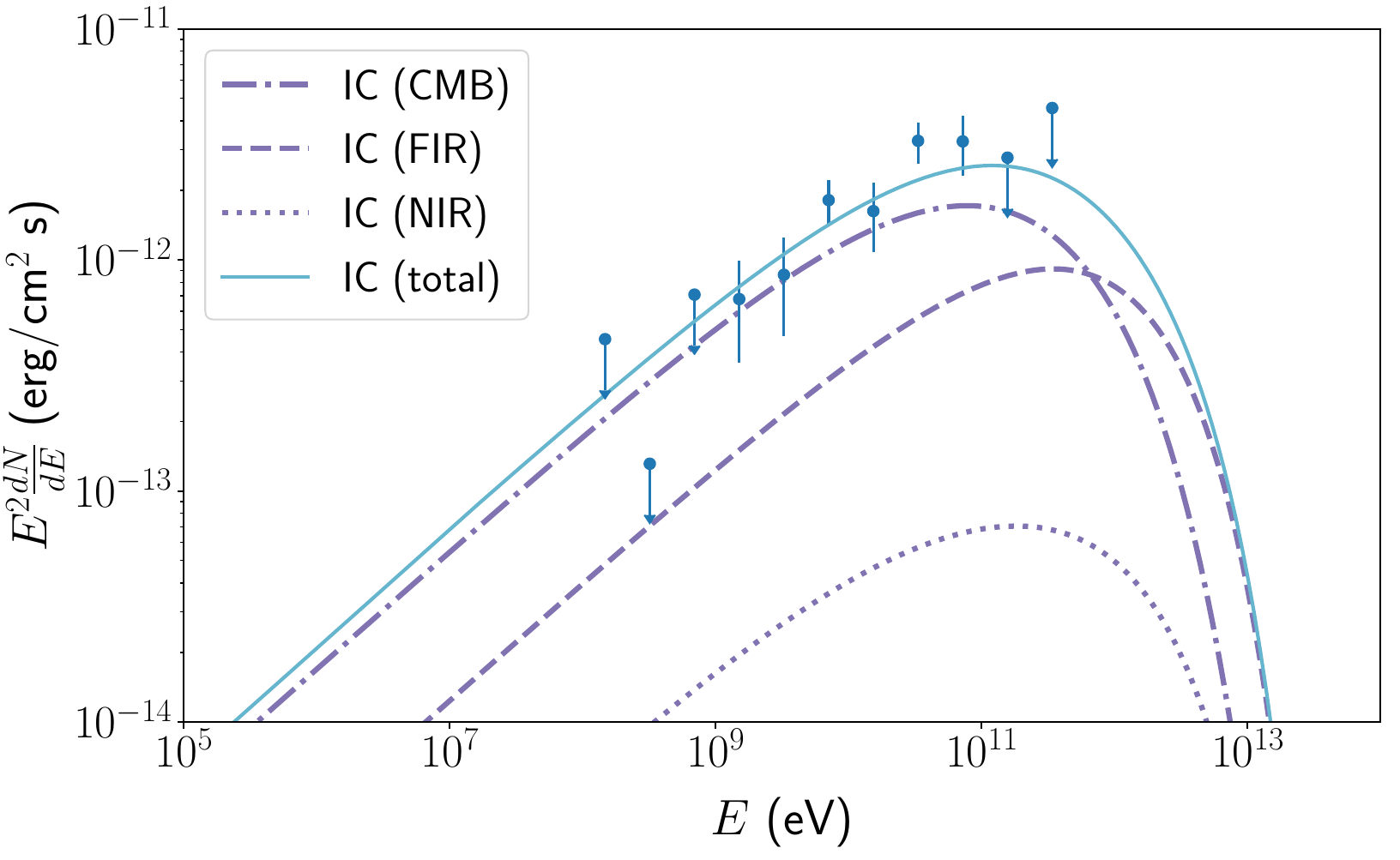}
\caption{A simple one-zone leptonic model for the non-thermal GeV emission from G$107.7-5.1$ analyzed in this work.
\label{fig3}}
\end{figure}

Fig. \ref{fig3} shows the GeV data from G$107.7-5.1$ and the leptonic model described in this section.

\section{Discussion}\label{discussion}
The gamma-ray emission presented in this work shows a morphology that is consistent with that of the newly-discovered SNR candidate G$107.7-5.1$ (the Nereides nebula), which we argue confirms the SNR nature of the object. It becomes a new example of an SNR at relatively high Galactic latitudes that is detected at GeV energies. Besides the seven objects listed by \cite{2024A&A...684A.150B}, other examples could include the Cygnus Loop, HB 21 (G$89.0+4.7$), G$332.5-5.6$ and the remnant of Kepler's supernova \citep{2011ApJ...741...44K,2013ApJ...779..179P,2021ApJ...908...22X,2024RAA....24d5012L}.

The spectral slope measured for G$107.7-5.1$ is hard ($\alpha \sim 1.67$). A softer spectral index was recently found for the Ancora SNR \citep[G$288.8-6.3$ with $\alpha \sim 2.3$,][]{2024A&A...684A.150B}, which might indicate that G$107.7-5.1$ is in a younger evolutionary stage. Examples of SNRs with a spectral slope at GeV energies similar to that of G$107.7-5.1$ are Calvera's SNR and G$150.3+4.5$ \citep{2023MNRAS.518.4132A,2020A&A...643A..28D}, both are radio-dim SNRs that likely expand in low-density environments, which could also be the case of the Nereides nebula. We note the GeV similarities between G$107.7-5.1$ and G$150.3+4.5$. Their photon spectral indices are very similar, and \cite{2020A&A...643A..28D} also found similar maximum electron energies of $\sim 5$~TeV in a leptonic (IC) scenario for the gamma rays.

As we showed in section \ref{leptonic} no molecular gas is seen in the region of the SNR. The WISE catalog of HII regions \citep{2014ApJS..212....1A} likewise shows no object close to the SNR. This would explain the low GeV luminosity of G$107.7-5.1$, $\sim 10^{33}$~erg~s$^{-1}$ (assuming a distance of 1~kpc) compared to those of SNRs interacting with molecular clouds \citep[e.g.,][]{2016ApJS..224....8A}.

We have shown a simple leptonic model that accounts for the GeV fluxes of the Nereides nebula. Although the particle spectral index of 2 found in the model is expected for a diffusive shock-accelerated electron spectrum, a relevant feature in the spectrum is its curvature, similar to that of G$150.3+4.5$ \citep{2020A&A...643A..28D}. This could be naturally explained by the cooling of the electrons responsible for the emission, which results in a spectral break. However, the estimated cooling times are relatively large. Assuming that electrons cool due to synchrotron and IC-CMB losses and this causes the observed particle spectral turnover at $\sim 4$~TeV, an age estimate can be obtained by calculating the corresponding cooling time as \citep{2006A&A...460..365A}, $$\tau_{\mbox{\tiny cool}} \approx \frac{77.5\,\,\mbox{kyr}}{0.26 + 0.025\,B^2_{\mu}},$$ with $B_{\mu}$ the magnetic field in units of $\mu$G. Using $B_\mu = 3$ the age is 160~kyr, which is not reasonable for an SNR with a well-defined shell. On the other hand for an average field value $B_\mu > 10$ the corresponding age is $<28$~kyr. We note that these estimates should be taken cautiously since the magnetic field is a function of time and was likely larger in the past making the particles cool faster. Future detailed observations in the radio and above $\sim 100$~GeV should confirm or rule out a spectral break and better constrain any spectral turnover.

We estimated the SNR age from the ambient parameters. From optical observations an angular radius of $\sim 1.2^\circ$ is determined for the SNR shell. The physical sizes of SNRs with well-defined shells rarely exceed 100~pc \citep[e.g.,][]{2010MNRAS.407.1301B}. Thus the distance to G$107.7-5.1$ is likely to be $d < 2.4$~kpc. Using a column density calculator employing the method by \cite{2013MNRAS.431..394W}\footnote{See \url{https://www.swift.ac.uk/analysis/nhtot/}.}, we found that the atomic hydrogen contribution dominates over that of molecular hydrogen in the direction of the SNR. With data from the HI4PI atomic hydrogen survey \citep[][spectral resolution $\sim1.49$~km~s$^{-1}$]{2016A&A...594A.116H} we chose four $\sim 5$~km~s$^{-1}$-wide local standard of rest (LSR) velocity ($v_{\mbox{\tiny lsr}}$) intervals and, assuming that the measured velocities arise from the differential Galactic rotation of the gas, we calculated the associated kinematic distances using the Galactic rotation model from \cite{1993A&A...275...67B}. We only selected velocities consistent with the constrain $d < 2.4$~kpc. Given the location of the SNR beyond the solar orbit, a value of $v_{\mbox{\tiny lsr}}$ uniquely determines the kinematic distance. In each velocity interval we integrated the brightness temperature of the atomic emission and estimated the column density ($N_H$) in the direction of the SNR in the optically thin limit \citep[see, e.g.,][]{1982AJ.....87..278D}. With an integration line-of-sight length of $\sim 0.5$~kpc for each interval and the average value of $N_H$ we estimated the atomic number density ($n$) of the interstellar medium. Finally, using the Sedov-Taylor (ST) solution we calculated the SNR age ($t_{\mbox{\tiny ST}}$) from its radius ($R_{\mbox{\tiny SNR}}$) and the ambient density, assuming an initial kinetic energy in the SNR of $10^{51}$~erg \citep[note that for the densities obtained here the SNR is predicted to still be in the ST stage, see][]{1988ApJ...334..252C}. The results are shown in Table \ref{table2}. The SNR age is constrained in the $10-50$~kyr range under the assumption that the ST solution applies.

\begin{table}
\begin{center}
\caption{SNR parameters.}
\label{table2}
\resizebox{\linewidth}{!}{
\begin{tabular}{cccccc}
\hline\hline
$v_{\mbox{\tiny lsr}}$ (km~s$^{-1}$) & $d$ (kpc) & $n$ (cm$^{-3}$)& $R_{\mbox{\tiny SNR}}$ (pc)& $t_{\mbox{\tiny ST}}$ (kyr) & $v_{\mbox{\tiny sh}}$ (km~s$^{-1}$)\\
\hline
($-20$, $-15$) & 2.2 & 0.04& 46& 52 & 346\\
($-15$, $-10$) & 1.75 & 0.1& 37& 48 & 302\\
($-10$, $-5$) & 1.25 & 0.2& 26& 28 & 363\\
($-5$, $0$) & 0.7 & 0.4 & 15 & 10 & 587\\
\hline
\end{tabular}}
\end{center}
\end{table}

For the different values of the SNR radii and ages we also estimated the ST shock speed ($v_{\mbox{\tiny sh}} = \dot{R}_{\mbox{\tiny SNR}}$), which is shown in Table \ref{table2}.

If the downstream magnetic field in the SNR has a similar value to that found for G$150.3+4.5$ ($\sim 5$~$\mu$G), the cooling time is $\sim 87$~kyr which is larger than the ST age. In this case, no break is expected in the particle distribution and a spectral turnover appears since the acceleration timescale is limited by the source age, $t_{\mbox{\tiny acc}}\approx t_{\mbox{\tiny ST}}$. The acceleration timescale is inversely proportional to the magnetic field and the square of the shock speed, and proportional to the maximum particle energy (here equal to 4~TeV) and the ratio between the mean free path and the gyroradius \citep{2006A&A...453..387P}, $k$. For evolved SNRs we expect that $k>1$. Adopting a shock compression ratio $r=4$ and the values of $v_{\mbox{\tiny sh}}$ with their corresponding ages $t_{\mbox{\tiny ST}}=52, 48, 28, 10$~kyr from Table \ref{table2} we obtain $k=177, 215, 87, 12$, respectively, which are plausible values. These estimates, however, assume a constant shock speed which is a considerable simplification.

If, on the other hand, the maximum electron energy observed for this source is the result of the balance of synchrotron losses and acceleration in a slow shock the condition $\tau_{\mbox{\tiny cool}}\approx t_{\mbox{\tiny acc}} < t_{\mbox{\tiny ST}}$ can be accomplished for the four cases $t_{\mbox{\tiny ST}}=52, 48, 28, 10$~kyr if $k\lesssim 3.6$ with $B\gtrsim 7\,\mu$G, $k\lesssim 2.6$ with $B\gtrsim 7.3\,\mu$G, $k\lesssim 3$ with $B\gtrsim 10\,\mu$G and $k\lesssim 4.8$ with $B\gtrsim 17\,\mu$G, respectively. These values for the parameters are also reasonable and this makes this scenario also possible.

In summary, the ST ages estimated are consistent with the GeV spectrum of G$107.7-5.1$.

Finally, although we note that the kinematic distance determination likely suffers from high uncertainties \citep{2018ApJ...856...52W} and the SNR parameters derived here should be taken cautiously, it seems likely that G$107.7-5.1$ is one more example in a population consisting of relatively evolved and dim SNRs detected at GeV energies, possibly evolving in low-density environments, and for which a leptonic scenario is a reasonable explanation for the origin of the gamma rays. The existence of such a population has been proposed before \citep{2020MNRAS.492.5980A} and some examples of similar objects are given by \cite{2023MNRAS.518.4132A}. Future, more detailed observations of the Nereides nebula should be carried out to better constrain the source properties and contribute to the understanding of SNR evolution.

\begin{acknowledgements}
We thank Robert A. Fesen for providing us with the optical image of SNR G$107.7-5.1$ and the anonymous referee for comments and suggestions that helped improve the quality of this work. We thank funding from Universidad de Costa Rica under grant number C4228.
\end{acknowledgements}

\bibliographystyle{aa} 
\bibliography{aanda.bib} 

\begin{appendix}
\section{Residual TS maps}\label{residuals}
\begin{figure}[h!]
\begin{minipage}{0.6\textwidth}
\begin{center}
\hspace*{-3.5cm}
\includegraphics[width=\textwidth]{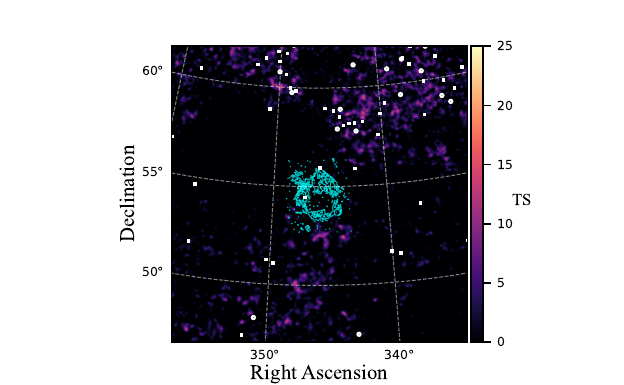}\\
\hspace*{-3.5cm}
(a)
\end{center}
\end{minipage}
\hfill
\begin{minipage}{0.6\textwidth}
\begin{center}
\hspace*{-3.5cm}
\includegraphics[width=\textwidth]{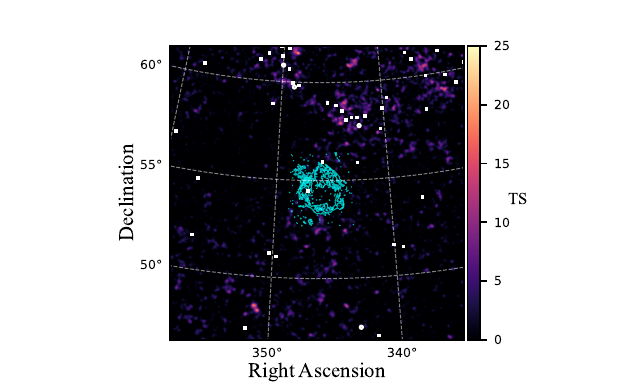}\\
\hspace*{-3.5cm}
(b)
\end{center}
\end{minipage}
\caption{Residual TS maps obtained after subtracting all the background 4FGL sources (squares), additional excess emission modeled as point sources in this work (circles) and the SNR G$107.7-5.1$, in two analysis intervals: 0.1--500~GeV (a) and 2--500~GeV (b). The contours represent the optical emission from the SNR shell.}
\label{fig:residuals}
\end{figure}

This section presents residual TS maps of a $15^\circ\times 15^\circ$-region around the location of G$107.7-5.1$ obtained after subtracting all the background sources as well as the emission from the SNR. Fig. \ref{fig:residuals} shows the maps for events with energies above 100~MeV and for events with energies above 2~GeV. The positions of the catalogued 4FGL sources and the new point sources added to the model during the optimization step are shown. The locations of the new point sources found above 2~GeV generally correspond with sources found at lower energies. However, more sources are found at lower energies, particularly towards the north of the SNR where the Galactic plane is located.


\FloatBarrier

\section{On the background sources 4FGL~J$2309.0+5425$ and 4FGL~J$2303.9+5554$}\label{10GeVmap}
\begin{figure}[h!]
\hspace*{-1cm}
\includegraphics[width=1.2\linewidth]{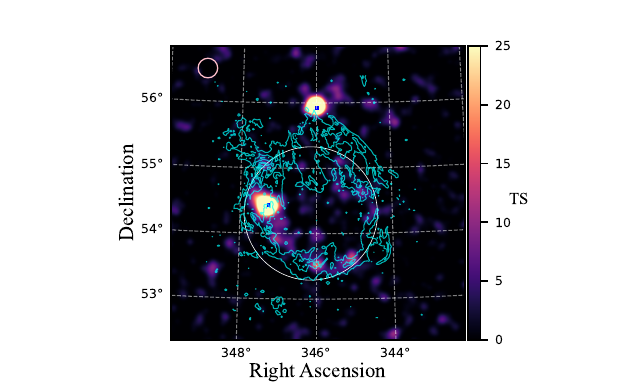}
\caption{A TS map above 10~GeV obtained after removing the SNR and the sources 4FGL~J$2309.0+5425$ and 4FGL~J$2303.9+5554$ from the model (thus causing them to appear in the map). The large circle indicates the gamma-ray disk model for the SNR (whose optical contours are shown) and the squares are the catalogued positions of the 4FGL sources. The upper-left circle indicates the 68\%-containment region of the LAT point spread function for both front- and back-converted events at 10~GeV.}
\label{10GeVPS}
\end{figure}

Fig. \ref{10GeVPS} is a TS map obtained with events having energies above 10~GeV showing the SNR emission as well as that of 4FGL~J$2309.0+5425$ and 4FGL~J$2303.9+5554$. The image clearly shows that these sources are point-like for the LAT and thus possibly associated to independent objects and not the shell of the SNR. As indicated previously, 4FGL~J$2309.0+5425$, seen in the eastern side of the SNR, is believed to be associated with an X-ray source, while 4FGL~J$2303.9+5554$, seen to the north was classified as a likely blazar by \cite{2021ApJ...923...75K}.

\FloatBarrier
\end{appendix}

\end{document}